# AI-Enhanced Resilience in Power Systems: Adversarial Deep Learning for Robust Short-Term Voltage Stability Assessment under Cyber-Attacks

Yang Li [1 *], Shitu Zhang [2], Yuanzheng Li [3]

1 School of Electrical Engineering, Northeast Electric Power University, Jilin, China

2 State Grid Hebei Provincial Electric Power Company Yuanshi County Power Supply Branch Company, Shijiazhuang, China

3 School of Artificial Intelligence and Automation, Huazhong University of Science and Technology, Wuhan, China

*Corresponding author. Yang Li (liyang@neepu.edu.cn)

**Abstract**: In the era of Industry 4.0, ensuring the resilience of cyber-physical systems against sophisticated cyber threats is increasingly critical. This study proposes a pioneering AI-based control framework that enhances short-term voltage stability assessments (STVSA) in power systems under complex composite cyber-attacks. First, by incorporating white-box and black-box adversarial attacks with Denial-of-Service (DoS) perturbations during training, composite adversarial attacks are implemented. Second, the application of Spectral Normalized Conditional Wasserstein Generative Adversarial Network with Gradient Penalty (SNCWGAN-GP) and Fast Gradient Sign Method (FGSM) strengthens the model's resistance to adversarial disturbances, improving data quality and training stability. Third, an assessment model based on Long Short-Term Memory (LSTM)-enhanced Graph Attention Network (L-GAT) is developed to capture dynamic relationships between the post-fault dynamic trajectories and electrical grid topology. Experimental results on the IEEE 39-bus test system demonstrate the efficacy and superiority of the proposed method in composite cyber-attack scenarios. This contribution is pivotal to advancing AI-based resilient control strategies for nonlinear dynamical systems, marking a substantial enhancement in the security of cyber-physical systems.

## 1 Introduction

With the continuous advancement of information technology, power systems are undergoing a period of digital and smart transformation. Although this process greatly enhances the efficiency and intelligence of power systems, it also brings more information and data, thereby increasing the possibility and complexity of cyber-attacks. The cybersecurity threats faced by power systems are becoming increasingly severe, with attackers potentially motivated by various reasons, such as personal gain or commercial competition. Several major cyber-attack incidents in history, such as the 2003 Slammer worm incident [1], the 2007 "Aurora" attack [2], and the recent blackout incidents in Ukraine [3], highlight the vulnerabilities in power system cybersecurity. These events have prompted governments worldwide to increase their focus on cybersecurity, leading to the introduction of a series of policies and measures to strengthen the cybersecurity of critical infrastructure [4,5].

**1.1 Advances and Limitations of STVSA**

Against the backdrop of increasingly digital and intelligent power systems, the development of Short-Term Voltage Stability Assessment (STVSA) methods has shown a diverse range of trends. These methods mainly include time-domain simulation [6], and data-driven artificial intelligence methods [7,8]. Each method has its unique advantages and application scenarios, but also certain limitations. Time-domain simulation, as a traditional method in power system analysis, is widely used in the industry for its excellent model adaptability and reliability [9]. This method solves Differential-Algebraic Equations (DAEs) through numerical integration to simulate the dynamic process of power system disturbances in detail, but it is computationally expensive, posing challenges for real-time or near-real-time applications [10]. Direct methods, utilizing energy functions based on Lyapunov theory, provide a means for quick assessment. For example, the Koopman model demonstrates that this method can effectively assess the stability of power systems without the need for complex time-domain simulations [11]. However, the accuracy and applicability of this method when dealing with complex power systems still require further research.

In recent years, with the rapid development of artificial intelligence technology, data-driven STVSA methods have gained widespread attention. Unlike traditional time-domain simulations and direct methods, data-driven methods do not require complex physical models. Instead, they achieve rapid assessment by learning the mapping relationship between power system operating parameters and stability states [12]. The advantage of this method lies in its rapid assessment speed and strong learning capability; however, it also faces challenges such as poor interpretability of results, strong dependence on data, and difficulties in model selection. To improve adaptability to topological changes in real-world scenarios, reference [7] enhances model generalization by employing deep transfer learning to explore the relationships among different fault types. This approach significantly improves model resilience during assessments. In the STVSA domain, reference [13] develops an intelligent machine learning method that employs deep recurrent neural networks with long short-term memory to capture the complex transient spatial and temporal dependencies within smart grids, leveraging both static network information and dynamic system responses. Similarly, reference [14] employs emerging graph convolutional networks to learn the non-Euclidean topological structure of power grids and uses recurrent long short-term memory algorithms to capture the temporal characteristics of short-term voltage stability (STVS) dynamics, developing a powerful structure-aware recurrent learning machine for precise prediction of voltage trajectory sensitivity indices. Further deepening the examination of STVS scenarios, reference [15] utilizes highly efficient spatio-temporal graph convolutional networks to derive diverse base learners for clustering, enhancing model robustness against data loss and achieving highly robust STVSA performance.

Unfortunately, despite the existence of several studies focusing on resilience scheduling for integrated energy systems [16] and research on single false data attacks in power systems [17], the investigation of model stability and robustness under multiple attacks in STVSA has not yet received sufficient attention. This is particularly true for classical Denial of Service (DoS) attacks and adversarial sample attacks originating from the field of image processing. These types of attacks pose serious challenges to the security of power systems. Therefore, it is crucial to enhance research in this area and explore effective defense mechanisms to ensure the stable operation and secure protection of power systems against a diverse range of threats [18]. Reference [19] proposes a mitigation strategy for false data attacks, incorporating power flow equations to handle vector attacks and using multi-port equivalent circuit calculations of Thevenin parameters as detection indicators. This approach enhances the resilience of cyber-physical systems against attacks. Reference [20] notes that data-driven models are susceptible to adversarial sample attacks during stability assessments, leading to severe model failures. Thus, it proposes a universal defense strategy based on smoothing random algorithms to enhance model resilience when under attack. Reference [21] analyzes the characteristics and applicability of different cyber-attack models, discussing decisive problem-solving approaches and defense mechanisms.

**1.2 Research Gaps of Existing Research**

Despite advancements in STVSA, significant challenges persist, particularly in tackling the complexities of modern power systems and escalating cybersecurity threats. Traditional methods struggle with processing large-scale data and lack adaptability to evolving cyber threats, compromising their effectiveness in real-world scenarios. While AI-based approaches have improved processing speeds and learning capabilities, they often fall short in ensuring robustness and reliability under cyber-attacks. Current research tends to focus on enhancing technical accuracy and computational efficiency, often overlooking the crucial aspect of resilience. This oversight has led to a notable research gap in maintaining model performance under diverse and sophisticated cyber threats. Moreover, while much of the current research on STVSA focuses on time-series data analysis, it frequently overlooks the impact of explicit topological relationships within power systems. Some methods, such as graph-based Transformer models, capture the network's topology through their multi-head self-attention mechanisms and utilize this data, but they neglect the temporal dynamic characteristics of power systems. For STVSA tasks in power systems with varied topologies and dynamic features, these methods are still underexplored and underdiscussed. These gaps reveal that existing STVSA methodologies do not adequately address the unpredictable nature of cyber threats, exposing vulnerabilities in system stability and security. This underscores the urgent need for more robust and adaptive solutions to enhance the resilience of power systems against cyber threats, which are crucial for ensuring the stability and reliability of power infrastructures amid an evolving threat landscape.

To address the challenges posed by cyber-attacks on STVSA, this paper proposes an AI-based control framework that enhances the robustness and accuracy of assessments. This paper achieves several key objectives: Firstly, it introduces and successfully deploys a hybrid model that integrates Long Short-Term Memory (LSTM) networks with Graph Attention Networks (GATs), significantly boosting the accuracy and robustness of STVSA. Secondly, we conducted a thorough analysis of the prevalent cyber-attacks in power systems and their impact on system stability. Through adversarial training, the resilience of the model against these attacks was greatly enhanced. Furthermore, we developed a highly efficient variant of the Conditional Wasserstein Generative Adversarial Network with Gradient Penalty (CWGAN-GP), called Spectral Normalized CWGAN-GP (SNCWGAN-GP), to generate diverse training data, thereby improving the model's generalizability and fortitude against unseen attacks. The unique contributions of this paper include:

1) **Hybrid Model Improvement:** This study designs an LSTM-enhanced Graph Attention Network (L-GAT) to capture the temporal dynamics of post-fault trajectories and their topological dependencies in power grids. By integrating sequential modeling (LSTM) with structural reasoning (GAT) through hierarchical attention mechanisms, this hybrid architecture improves robustness of STVSA against composite cyber-attacks.

2) **System Resilience Enhancement:** This work innovatively investigates STVSA against composite cyber-attacks by combining white-box and black-box adversarial attacks with DoS perturbations. Exposing the model to both attack types during training enhances its robustness and generalization across cyber-attack scenarios. Through adversarial deep learning, the model's resilience is significantly strengthened, demonstrating its effectiveness in handling complex cyber threats.

3) **Data Augmentation Techniques Application:** This study develops SNCWGAN-GP, a spectral-normalized variant of CWGAN-GP, which enforces Lipschitz continuity via discriminator weight normalization to stabilize adversarial training and accelerate convergence. By generating synthetic data covering rare grid faults, the method improves the model's generalization ability and strengthens its robustness against unseen adversarial attacks.

## 2 Modeling and Implementation of Cyber-Attacks

Power systems are critical infrastructure, significantly impacting national stability, the economy, and individual livelihoods. In this era of rapidly advancing information technology, the level of digitization within power systems is increasing, elevating the importance of cybersecurity. In this paper, we utilize Phase Measurement Units (PMUs) technology rather than traditional remote terminal unit-based measurement techniques. PMUs offer high sampling frequencies and accuracy, enabling the acquisition of precise power system state data, which is crucial for STVSA. However, the detailed data collection also poses risks, as attackers could tamper with PMU data to disrupt the stable operation of the power system, potentially leading to large-scale failures.

STVSA is a vital element of power system stability analysis, addressing time scales from tens of milliseconds to several seconds and primarily focusing on the transient voltage processes caused by major disturbances. During these events, the dynamic characteristics of various system devices—such as protection devices, governors, exciters, and voltage regulators—play significant roles in system stability. If these devices are compromised by cyber-attacks, their altered dynamic characteristics can destabilize the power system and potentially lead to widespread blackouts. Fig. 1 illustrates the basic structure of a power system.

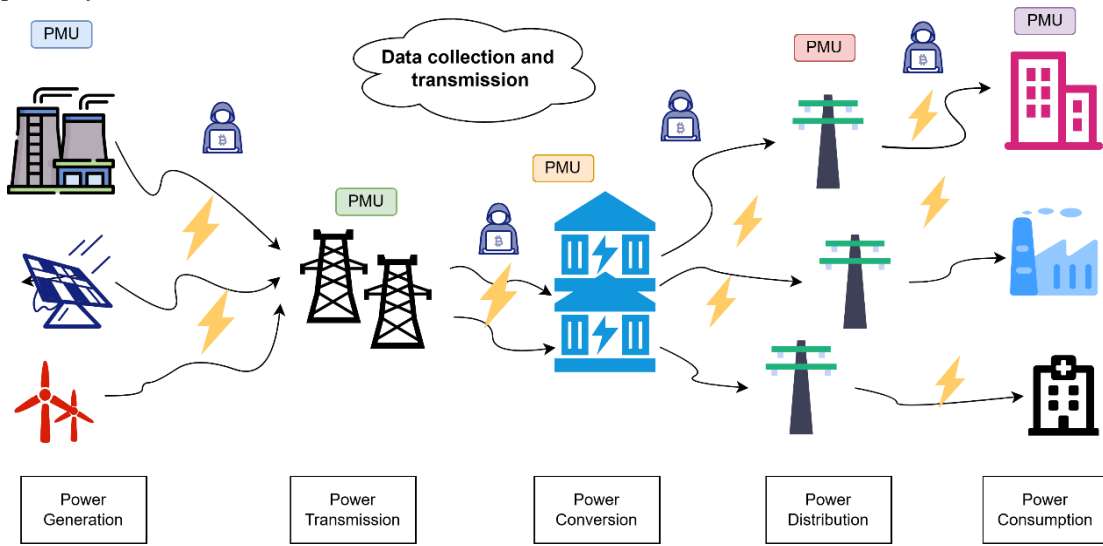

**Fig. 1.** Basic structure of a power system

As illustrated in Fig. 1, the power system includes generation, transmission, distribution, and consumption stages, each susceptible to cyber-attacks. PMUs are extensively deployed across these stages, and their data is critical for system stability. The interconnected nature of these stages means that cyber-attacks can set off cascading failures, inflicting significant damage. Therefore, understanding and mitigating cyber-attacks on power systems is vital for maintaining their stable operation and ensuring societal functionality [22]. Research in this area is essential for understanding how to prevent such attacks and maintain the stability of power systems.

**2.1 Adversarial Sample Attacks**

Adversarial sample attacks exploit the inherent vulnerabilities in machine learning models by introducing precise, usually imperceptible, perturbations to the inputs, leading the models to make incorrect decisions. These attacks are notably effective against deep learning models, revealing critical flaws in their data processing capabilities. At a broader level, adversarial attacks highlight the limitations of artificial intelligence in processing complex data structures. Investigating these attacks is imperative for identifying and fortifying the vulnerabilities of models against security threats, as well as for enhancing our understanding of the decision-making blind spots within these models. This research is crucial in areas like autonomous vehicles, facial recognition systems, and cybersecurity [23], where adversarial threats can pose severe security risks. Thus, studying adversarial sample attacks is fundamental to enhancing model robustness. It not only fosters advancements in security mechanisms and model design optimization but also encourages a deeper investigation into the models' internal workings. In scientific and engineering contexts, this line of inquiry prompts a re-evaluation of core algorithmic principles and the development of innovative defense strategies to mitigate future threats.

**2.1.1 Fast Gradient Sign Method**

In today's networked environments, the diversity of adversarial sample attacks is expanding, with each type showcasing unique characteristics and specific applications. Prominent among current mainstream adversarial attack algorithms are the Fast Gradient Sign Method (FGSM), Projected Gradient Descent (PGD), DeepFool, and the Carlini & Wagner Attack (C&W Attack) [24]. PGD builds on FGSM by incorporating iterative optimization of adversarial perturbations, which enhances attack precision and stability in convergence. DeepFool utilizes geometric approximations to calculate the minimal-norm perturbations necessary for misclassification, emphasizing the efficiency of creating imperceptible adversarial examples. The C&W Attack formulates adversarial generation as an optimization problem with a customized loss function, achieving near-optimal attack success rates at the expense of increased computational complexity.

FGSM is favored in both academic and industry settings due to its operational simplicity and rapid execution. The principal mechanism of FGSM involves calculating the gradient of the model's loss function relative to the input data, applying a sign function to this gradient, and then multiplying it by a small predetermined perturbation factor, ε. This perturbation is added to the original input to produce adversarial samples. This method intuitively demonstrates how to maximize the model's loss by subtly adjusting each input feature, thereby compelling the model to make incorrect decisions when faced with strategically crafted perturbations.

A notable advantage of FGSM is its speed, enabling the rapid generation of adversarial samples. This feature is particularly valuable in scenarios that demand quick reactions, such as in real-time systems and extensive model evaluations. In the context of STVSA for power systems, employing FGSM to simulate cyber-attacks offers substantial benefits. The method's simplicity, ease of implementation, and swift action are effective in promptly producing adversarial samples in time-sensitive environments within power systems. This capability allows system operators to quickly anticipate and mitigate potential cyber threats, boosting the system's security. The subsequent sections detail the algorithm's formulation and principles [25].

During the STVSA of power systems, given the paired initial training sample $\mathbf{x}=[x_t, t=1,\ldots,T]$ and label $y$, where $T$ represents the sampling moment, the learning function of the data-driven model based on machine learning is $f_\theta(\cdot)$ as shown in the equation, mapping $\mathbf{x}$ to $y$ with model parameters $\theta$. Here, $\mathbf{x}$ represents voltage trajectory values, and $y$ represents the corresponding stability status.

$$f_\theta(\mathbf{x}) = f^{(m)}\left(\ldots f^{(2)}\left(f^{(1)}(\mathbf{x})\right)\right) \tag{1}$$

where $f^{(h)}$ denotes the function of the $h$ layer of the neural network, $h=1,2,\ldots m$. In machine learning model training, the goal is to minimize the difference between the predicted $f_\theta(\mathbf{x})$ and the ground truth label $y$, which is formulated by

$$\min_\theta L_f(f_\theta(x), y) \tag{2}$$

where $L(\cdot,\cdot)$ is the predefined model loss function $f_\theta(\cdot)$, typically solved in neural networks using the gradient descent algorithm in a backpropagation process to update model parameters, i.e.

$$\theta_{i+1} = \theta_i - \eta \cdot \nabla_\theta L_f(f_\theta(\mathbf{x}), y) \tag{3}$$

where $\eta$ is the learning rate, $i$ denotes the $i$-th iteration step, $\nabla_\theta$ represents the partial derivative, and $\theta_i$ denotes the model parameters. Once the training sample $\mathbf{x}$ and label $y$ are available, accurate machine learning models can be obtained through various neural network training algorithms. Unlike the machine learning model training process, adversarial example generation strategy is based on an already trained machine learning model with parameters $\theta$, aiming to mislead the model. Given a trained classifier $f_\theta(\cdot)$, a sample $\mathbf{x}=[x_t, t=1,\ldots,T]$, and its ground truth label $y$, an adversarial example $\mathbf{x}^{adv}=[x_t^{adv}, t=1,\ldots,T]$ can be generated by solving the following optimization problem:

$$\min_{\mathbf{x}^{adv}} \quad \|x^{adv} - x\|_p \\ \text{s.t.:} \quad \begin{cases} f_\theta(x) = y \\ f_\theta(x^{adv}) = y^{adv} \neq y \end{cases} \tag{4}$$

where $y$ and $y^{adv}$ represent the output labels corresponding to $\mathbf{x}$ and $\mathbf{x}^{adv}$; $\|\cdot\|_p$ denotes the distance between $\mathbf{x}$ and $\mathbf{x}^{adv}$; $p$ indicates the $p$-norm distance representing the magnitude of the adversarial perturbation. Using gradient-based algorithms, adversarial examples $\mathbf{x}^{adv}$ can be quickly computed. FGSM, as a popular fast method, updates the gradient in the direction of the gradient's sign in one step, as shown below:

$$\mathbf{x}^{adv} = \mathbf{x} + \delta \cdot \text{sign}\left(\nabla_\mathbf{x} L(f_\theta(\mathbf{x}), y)\right) \tag{5}$$

where $\text{sign}(\cdot)$ is the sign function, $\delta$ specifies the perturbation boundary.

**2.1.2 Adversarial Sample Attack Generation Strategy**

This section outlines the strategy for generating adversarial sample attacks. Initially, we collect a dataset to train the deep learning model. This dataset, crucial for the training process, is compiled through time-domain simulations and further processed to enhance its diversity and volume, ensuring the adversarial samples generated possess robust generalization capabilities. For generating these adversarial samples, this paper employs the FGSM. This method utilizes backward gradient propagation to calculate the gradient of the loss function. It then introduces slight perturbations to the original data to maximize the loss function, effectively making the adversarial samples sufficiently similar to the original dataset to deceive the model. Following initial generation, the adversarial samples undergo a rigorous evaluation process. Based on these results, we fine-tune the attack strategy and continuously regenerate adversarial samples until the desired efficacy of the attack is achieved. The iterative process of generating and refining these samples is illustrated in Fig. 2.

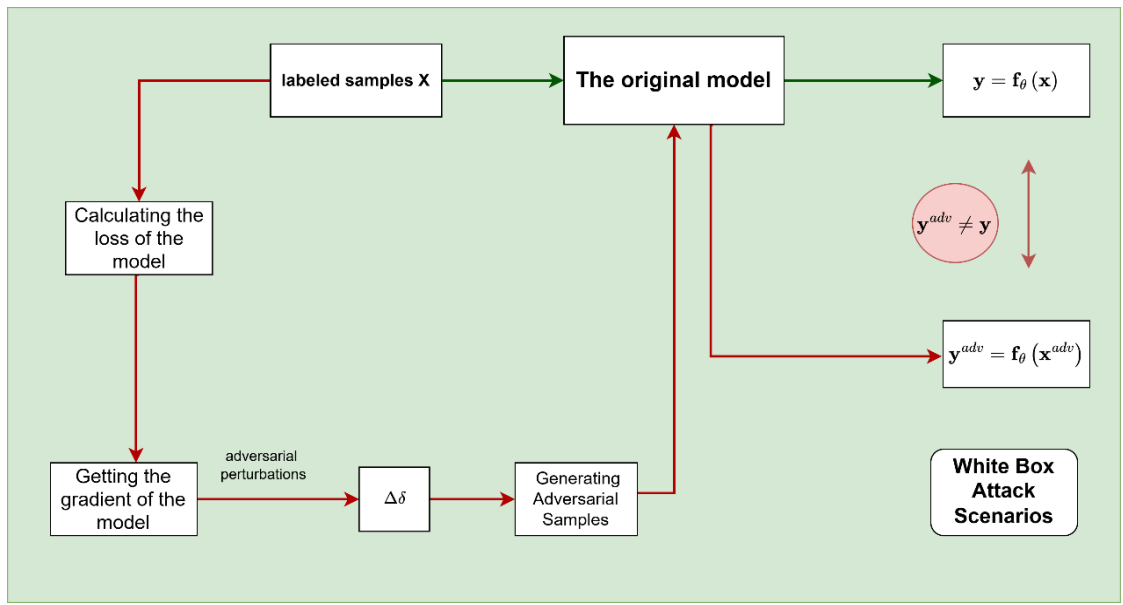

**Fig. 2 (a)** Process of generating adversarial examples in a white-box scenario

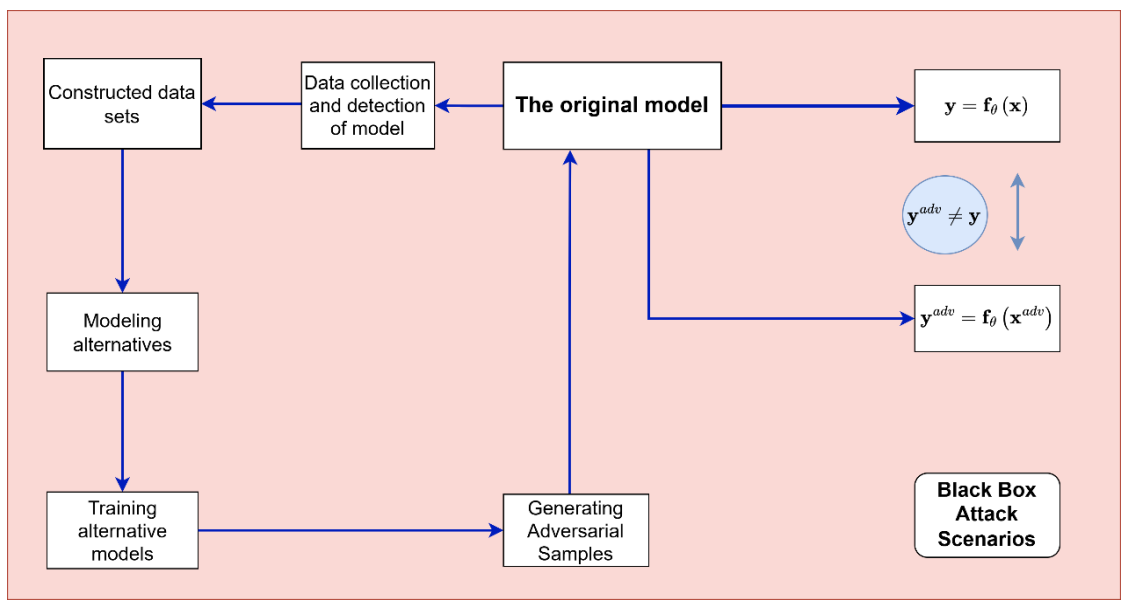

**Fig. 2 (b)** Process of generating adversarial examples in a black-box scenario

For White-Box Attack Scenarios (see Fig. 2(a)): he process begins with the initialization of the original deep learning model, input data $X$, original label $y$, target label $y^{adv}$, and hyperparameters. A loop operation is then executed where, in each iteration, the model's gradient is reset to zero. The loss between the model's output for the input data and the target label is calculated, and this loss is backpropagated to determine the gradient. An adversarial perturbation, derived from this gradient and a predefined small perturbation intensity, is added to the input data to generate an adversarial sample $X^{adv}$. The model then predicts the label $y^{adv}$ for the adversarial sample. If the predicted label is closer to the target label and deviates from the original label, the loop terminates. Otherwise, the input data for the next loop is replaced with the adversarial sample generated in the current iteration. The loop concludes with the generation of the final adversarial sample $X^{adv}$.

For Black-Box Attack Scenarios (as illustrated in Fig. 2(b)): In black-box scenarios, attackers lack knowledge about the internal structure and parameters of the model. Here, the application of FGSM typically relies on attacks against similar models or on methods that indirectly obtain model gradients or classification data. This usually involves training a simple surrogate model on existing data to approximate the target model's behavior, which is then used to generate adversarial samples. While generally less effective than white-box attacks, this approach can still assess the model's robustness and is more indicative of real-world scenarios that power systems might encounter. It is important to note that although FGSM is a popular attack method, the adversarial examples it produces might not always disrupt as intended, necessitating further iterations to refine and produce effective attack samples.

**2.2 Denial of Service Attacks**

DoS attacks constitute a widespread type of network assault, primarily aimed at disrupting network services. Within the realm of power systems, these attacks target the network communication infrastructure of power control systems. By flooding these systems with excessive requests or data packets, DoS attacks hinder the normal operations of power monitoring and control systems. This interference disrupts the flow of operational data across the power grid, significantly impacting real-time monitoring and decision-making processes essential for effective grid management. Such disruptions compromise the stability and security of the power supply. DoS attacks often focus on critical infrastructure components within power systems, including smart substations, data centers, and communication networks. These assaults can lead to diminished responsiveness or complete failure of automated equipment within power systems. In severe instances, this could escalate to widespread power outages.

In the field of power system security research, the Poisson distribution is commonly used to model the frequency and impact of DoS attacks [26]. This statistical approach helps simulate the occurrence of independent random events, such as DoS attacks, within a specified time frame. The Poisson distribution is crucial for analyzing count data, providing a mathematical framework to understand the probabilistic nature of these attacks. The formula for the Poisson distribution is given as follows:

$$p(k) = \frac{e^{-\lambda} \lambda^k}{k!} \qquad (6)$$

Here, $\lambda$ represents the average number of expected events per unit time, $k$ denotes the number of events that might be observed, and $e$ stands for the base of the natural logarithm. This formula calculates the probability of observing exactly $k$ attacks within a given timeframe. By adjusting the $\lambda$ parameter, ones can model attacks of varying intensities, helping to assess their potential impact on the security of power systems.

To simulate the effects of DoS attacks under severe scenarios, this study employs a technique that zeroes out parts of the data samples based on probabilities derived from the Poisson distribution. This method directly manipulates actual data to reflect the consequences of the attacks, providing a robust mechanism for analyzing the cybersecurity threats to power systems. Such simulations are instrumental in gaining a deeper understanding of the potential damage DoS attacks can inflict in real-world settings and lay a crucial foundation for developing more effective control strategies against these cyber threats.

**2.3 Composite Adversarial Attacks**

To simulate realistic cybersecurity threats on STVSA models, we extend conventional adversarial attacks by integrating them with DoS attacks, forming composite adversarial attacks. By combining adversarial and DoS perturbations, we emulate the worst-case scenarios envisioned in critical infrastructure standards. In this study, composite adversarial attacks (CAAs) are categorized into two subtypes:

- White-box composite adversarial attacks (WB-CAAs): Combining white-box adversarial attacks with DoS attacks.
- Black-box composite adversarial attacks (BB-CAAs): Integrating black-box adversarial attacks with DoS attacks.

For a composite attack on input $x$:

$$\mathbf{x}^{adv+DoS} = \underbrace{\mathbf{x} + \delta \cdot \text{sign}\left(\nabla_{\mathbf{x}} L(f_\theta(\mathbf{x}), y)\right)}_{\text{FGSM perturbation (Eq. (5))}} \odot \underbrace{M^{DoS}}_{\text{DoS mask (Eq. (6))}} \quad (7)$$

where $M^{DoS}$ is a binary ask nullifying data segments with probability $p(k)$.

# 3 Data Clustering and Augmentation

In STVSA, one fundamental challenge with data-driven methods is generating labeled samples and acquiring large-scale data. Given the highly nonlinear and dynamic characteristics of power systems, comprehensive and accurate data collection is particularly demanding. Moreover, the use of relevant data is often constrained by privacy concerns, commercial sensitivity, or technical limitations. As a result, generating basic and diverse experimental data through simulation software emerges as a promising approach to overcome these data acquisition challenges.

However, the training of AI and deep learning models on small datasets derived from simulations presents issues such as limited data scale, data homogenization, and category imbalance. These challenges significantly hinder model performance and restrict the potential development and application of AI-based models. To address these issues, this paper adopts SNCWGAN-GP-based data augmentation techniques to effectively process small-scale, imbalanced datasets from simulations. The goal is to train an AI model on a large-scale, balanced dataset that demonstrates superior performance and robustness.

**3.1 Semi-supervised Fuzzy C-Means Clustering**

In the domain of STVSA, the absence of a unified standard for stability determination complicates the labeling of a large number of samples as stable or unstable under diverse operational conditions of power systems. Given the high costs and logistical challenges associated with labeling extensive data volumes, semi-supervised clustering learning offers an innovative solution. This method combines features of both unsupervised and supervised learning, making it ideal for clustering tasks where only a subset of the data is labeled, while the majority remains unlabeled. Semi-supervised clustering employs a limited amount of labeled data to guide the clustering process, significantly enhancing task performance. It bridges the gap between purely unsupervised clustering and fully supervised classification tasks, markedly improving the accuracy and reliability of dataset clustering.

The Semi-supervised Fuzzy C-Means (SFCM) clustering algorithm extends the traditional Fuzzy C-Means (FCM) by incorporating elements of supervised learning and fuzzy logic into the clustering process. This enhancement improves the precision and reliability of clustering outcomes by utilizing fuzzy membership relations and supervised information to more accurately segment the dataset. In SFCM, the objective function is adapted to include supervisory information beyond what is used in traditional FCM, often through penalty terms that reinforce the influence of the labeled data on the clustering process. The objective function of SFCM is formulated as follows:

$$\min J = \sum_{i=1}^{N}\sum_{j=1}^{C} u_{ij}^2 \| x_i - c_j \|^2 + \lambda \sum_{i=1}^{N}\sum_{j=1}^{C} (u_{ij} - f_{ij}b_j)^2 \| x_i - c_j \|^2 \tag{8}$$

where $u_{ij}$ represents the membership degree of data point $x_i$ to cluster center $c_j$; $x_i$ refers to an individual data point within the dataset; $c_j$ indicates a cluster center; $N$ specifies the total count of data points; $C$ denotes the number of clusters; $\|x_i - c_j\|$ indicates the Euclidean distance between the data point $x_i$ and the cluster center $c_j$; $\lambda$ represents a regularization factor that ensures a balance between the compactness of the clusters and the adherence to supervisory information; $f_{ij}$ is the affiliation matrix of the labeled sample; and $b_j$ is a boolean two-valued vector. In this case, the affiliation update formula is described as:

$$u_{ij} = \frac{1}{1+\lambda} \left[ \frac{1 + \lambda\left(1 - b_j \sum_{i=1}^{N} f_{ij}\right)}{\sum_{k=1}^{N}\left(\frac{\|x_i - c_j\|}{\|x_k - c_j\|}\right)^2} + \lambda f_{ij}b_j \right] \tag{9}$$

This adjusts the membership degree of each data point $x_i$ to cluster center $c_j$ by incorporating distances to all cluster centers. The membership degrees are recalculated iteratively to minimize the objective function, which considers the distances between data points and cluster centers. The process of updating the cluster centers is another essential component, formally defined by:

$$c_j = \frac{\sum_{j=1}^{C} u_{ij}^2 x_i}{\sum_{j=1}^{C} u_{ij}^2} \tag{10}$$

This equation calculates each centroid $c_j$ as the weighted mean of all data points. The cluster centers are updated based on these new calculations to better represent the centroids of their respective clusters.

**3.2 Advantages of SNCWGAN-GP**

The Least Squares Generative Adversarial Network (LSGAN) [24, 25] represents a significant improvement over the traditional Generative Adversarial Network (GAN) by substituting the original cross-entropy loss function with a least squares loss function. This modification addresses some fundamental issues in GAN training, notably enhancing the stability of network training and facilitating model convergence. LSGAN has demonstrated substantial potential in fields such as image processing and data augmentation, where high-quality data generation is critical. Despite these advancements,

LSGAN encounters challenges when applied to data from complex systems, including issues with poor convergence, homogenization of generated samples, and incomplete data distribution. To overcome these limitations, this paper extends the Conditional Wasserstein GAN with Gradient Penalty (CWGAN-GP) framework by integrating constraints and calculations into the discriminator's weight matrix, leading to the development of the Spectrally Normalized Conditional Wasserstein GAN with Gradient Penalty (SNCWGAN-GP) model. This enhanced model utilizes the Wasserstein loss, characteristic of all GANs based on the Wasserstein distance, which complies with the 1-Lipschitz condition. This compliance ensures the theoretical assurances of CWGAN-GP are met, substantially improving the efficiency of model training. Moreover, it effectively mitigates common issues such as gradient explosion or vanishing during training, thereby boosting the stability of the discriminator throughout the training process. Ultimately, these improvements enable the generation of higher-quality data.

A pivotal enhancement in the SNCWGAN-GP model is the incorporation of spectral normalization into the standard CWGAN-GP architecture. This modification aims to stabilize the training process and ensure that the discriminator strictly adheres to the Lipschitz constraints. Here are the primary mathematical expressions and formulas used in the SNCWGAN-GP model, along with detailed explanations: The Wasserstein Loss, calculated between the generator $G$ and the discriminator $D$, addresses the mode collapse issue commonly observed in traditional GAN training. The loss function is calculated as follows:

$$\min_G \max_D V(D,G) = \mathbb{E}_{x \sim p_{\text{data}}(x)}[D(x|y)] - \mathbb{E}_{\tilde{x} \sim p_{\tilde{x}}}[D(\tilde{x}|y)] - \lambda \mathbb{E}_{\hat{x} \sim p_{\hat{x}}}\left[(\|\nabla_{\hat{x}} D(\hat{x}|y)\|_2 - 1)^2\right] \quad (11)$$

where $x$ denotes samples from the real data distribution; $p_{\text{data}}(x)$ represents the distribution of real examples; $p_z(x)$ is a noise distribution; and $z$ indicates the input noise vector for the generator. $D$ is the discriminator; $p_{\hat{x}}$ represents the linearly interpolated distribution between the real examples' distribution $p_{\text{data}}(x)$ and the generated examples' distribution $p_{\tilde{x}}$, with $\hat{x} = \varepsilon x + (1-\varepsilon)\tilde{x}$ where $\varepsilon$ is a uniformly sampled random number from [0, 1] and $\tilde{x}$ stands for the generated examples from $G(z)$; the gradient penalty coefficient $\lambda$ is set to 10; and $y$ represents the additional auxiliary information, which in this approach refers to the class label. To ensure that the discriminator complies with the Lipschitz constraint, the SNCWGAN-GP incorporates a gradient penalty. This penalty is an additional term in the loss function designed to restrain the magnitude of the discriminator's gradients. The gradient penalty is defined as:

$$L_{\text{GP}} = \lambda \cdot \mathbb{E}_{\hat{x} \sim P_{\hat{x}}}\left[(\|\nabla_{\hat{x}} D(\hat{x}|y)\|_2 - 1)^2\right] \quad (12)$$

where $\hat{x}$ is a mixture of real and generated samples, typically obtained through linear interpolation; $\lambda$ is the coefficient of the gradient penalty; $\nabla_{\hat{x}} D(\hat{x})$ denotes the gradient of the discriminator $D$ evaluated at $\hat{x}$. Spectral normalization is applied to each layer of the discriminator $D$ to ensure that these layers conform to a 1-Lipschitz function. The spectral normalization formula is:

$$W_{\text{sn}} = \frac{W}{\sigma(W)} \quad (13)$$

where $W_{sn}$ denotes the normalized weight matrix; $W$ is the unnormalized weight matrix; and $\sigma(W)$ represents the largest singular value of $W$. Spectral normalization limits the spectral norm of the weight matrices, contributing to the stabilization of the CWGAN-GP training process. The objective function of SNCWGAN-GP directly uses the score output by the discriminator $D$, instead of its log probability, to calculate the difference between the two sets of data. This formula implies that the discriminator attempts to maximize the difference between the real and generated data, thereby enhancing the discriminator's ability to distinguish between the two. The objective function for SNCWGAN-GP is formulated as:

$$\min_G \max_D \left[ \mathbb{E}_{x,y}[D(x|y)] - \mathbb{E}_{z,y}[D(G(z|y)|y)] - \lambda \mathbb{E}_{\hat{x},y}\left[ \left( \|\nabla_{\hat{x}} D(\hat{x}|y)\|_2 - 1 \right)^2 \right] \right] \quad (14)$$

The combination of these key design features makes the SNCWGAN-GP excel in generating high-quality data, and in particular shows strong capabilities in image generation and sample generation for high-dimensional complex systems.

**3.3 Principle and Structure of the SNCWGAN-GP**

Fig. 3 illustrates the workflow of the SNCWGAN-GP along with a schematic diagram outlining the principles of its key components.

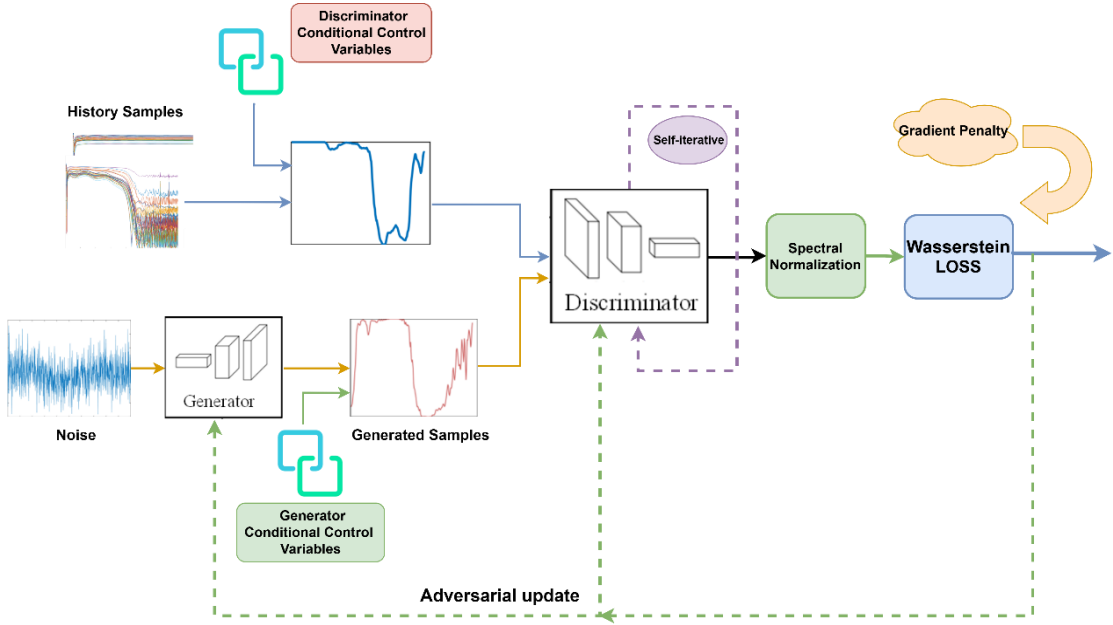

**Fig. 3.** Schematic diagram of the SNCWGAN-GP

The SNCWGAN-GP consists mainly of two parts: the generator and the discriminator, both of which are influenced by controlled variables. Under these constraints, the generator leverages noise to produce diverse and high-quality synthetic samples. These samples, combined with real data, are fed into the discriminator, which also responds to changes in the controlled variables and may require frequent self-updates to adapt to evolving data structures. A critical aspect of these updates is the normalization of the output weight matrix, constrained by spectral norms. Feedback from the discriminator's loss function is used to adjust the generator's parameters through a gradient penalty mechanism, optimizing the Wasserstein loss to refine the model's capability to generate realistic data. This process is iterative, promoting continuous enhancements in the quality of the generated data as depicted in the self-iterative cycle shown in the diagram.

Fig. 4 details the basic structure of the generator and discriminator in the SNCWGAN-GP during the data processing phase. The generator includes a label embedding layer, transposed convolution layers, activation layers, and an output layer. The label embedding layer is pivotal, transforming category labels into dense vector representations, which is crucial for conditional GANs as it allows the model to consider category-specific attributes during the discrimination process. Transposed convolution layers incrementally increase the dimensionality of the data, which is typical in generator architectures and helps the generator learn to produce spatially structured data from latent space vectors. Activation layers introduce nonlinearity, enabling the network to learn complex functions and thus generate more sophisticated and high-quality data. The output layer finalizes the data synthesis by delivering the generated output.

The discriminator comprises an input layer, convolutional layers, spectral normalization, a fully connected layer, and an output layer. The input layer receives external data, typically a mix of real and synthetic samples. Convolutional layers hierarchically extract features from this input, enhancing the model's sensitivity to subtle data variations, essential for differentiating between real and generated data. Spectral normalization within the convolutional layers curbs potential gradient explosions during weight updates, maintaining training stability. The fully connected layer synthesizes the processed features into a final discriminative output, assessing the authenticity of the data. The output layer then conveys the discrimination results, providing crucial feedback for training the generator and helping to refine the generation strategy for more realistic outputs.

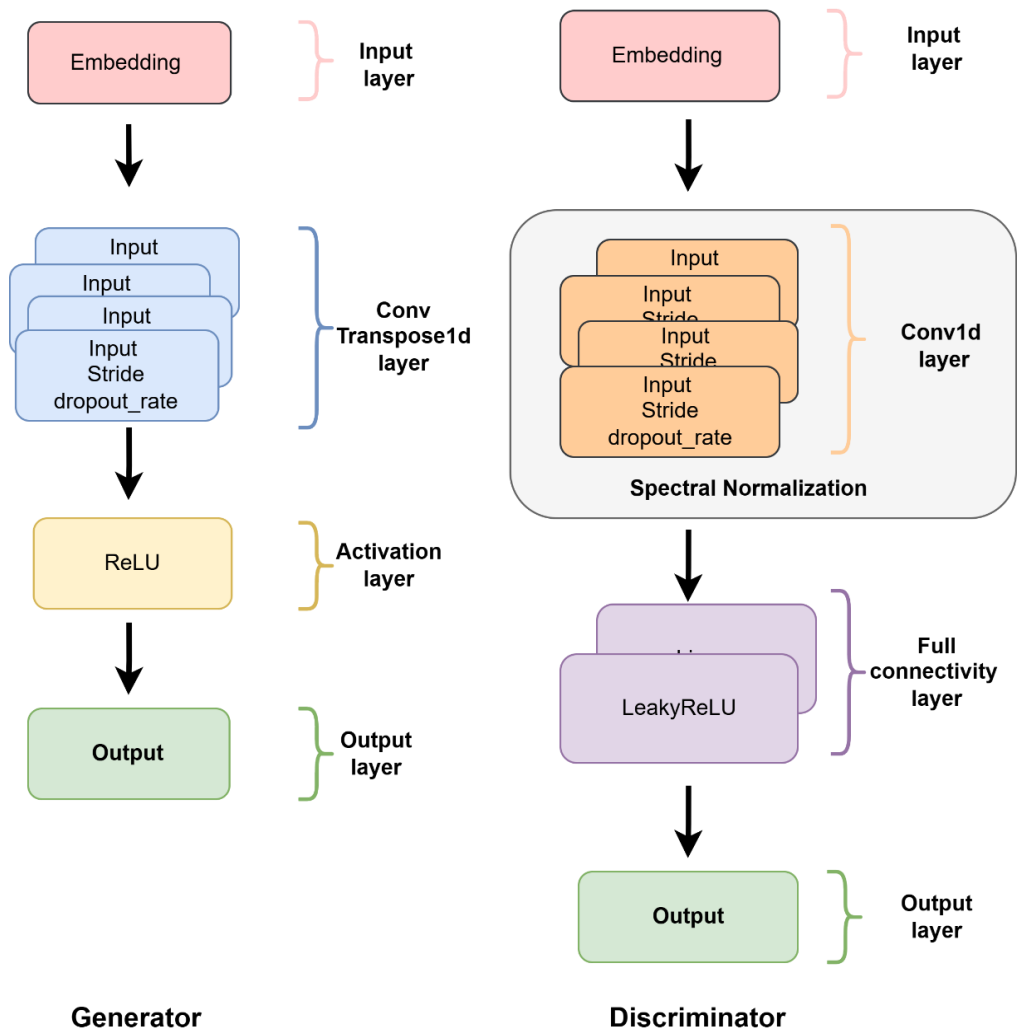

**Fig. 4.** Structure of the SNCWGAN-GP

# 4 LSTM-enhanced Graph Attention Network

In the field of power system security assessment, the adoption of Graph Convolutional Networks (GCNs) and Transformer models for system state classification and pattern prediction has recently become a prominent area of research. However, these advanced methodologies still encounter significant challenges in practical applications. Firstly, traditional graph neural networks like GCNs depend on a static graph structure and aggregate features from neighboring nodes via the adjacency matrix. This aggregation often involves simple averaging or weighted averaging, assuming uniform influence from all adjacent nodes, thus neglecting potential critical differences among them. Secondly, while traditional Transformer models have shown considerable success in Natural Language Processing, they were not inherently designed to handle graph-structured data. Transformers mainly utilize positional encoding to interpret input data, disregarding the connectivity or the local structural characteristics of the nodes, which reduces their adaptability to networks with varied topologies.

To address these limitations, recent research has explored integrating graph neural networks with attention mechanisms, culminating in the development of GATs. GAT dynamically assigns weights to adjacent nodes and learns the network topology, substantially improving the model's handling of graph data. Nevertheless, despite GAT's strengths in weight allocation and topology learning, it has not adequately addressed the dynamic changes in system states over time, especially in time-sensitive domains like power systems, thereby limiting its effectiveness.

## 4.1 Basic Principles of LSTM-enhanced Graph Attention Networks

This study advances the GAT model by incorporating a LSTM network, creating an augmented model termed L-GAT. This model is specifically engineered to more effectively capture the dynamic changes in node features over time. In L-GAT, time-series data from each node is independently processed through an LSTM layer, allowing the model to discern temporal dependencies and critical pattern features, thus enhancing its responsiveness to time-series data. This integration not only preserves GAT's benefits in learning topologies and allocating weights among neighbors but also harnesses LSTM's prowess in temporal modeling to refine the model's perception of evolving system states.

While the L-GAT model incorporates the attention mechanism typical of Transformers within the framework of graph neural networks—especially in terms of dynamically modulating interaction strengths between nodes based on data changes—it does not fully adopt the complete architecture of traditional Transformer models. Instead, L-GAT adapts its graph attention layers to better process graph-structured data, differing from the classical self-attention mechanism of traditional Transformers, which primarily focuses on node-to-node relationships.

## 4.2 Core Mechanisms of L-GAT

The functionalities of the L-GAT are elaborately depicted in Fig. 5, showcasing how the model employs its graph attention layers to master and execute complex interactions between nodes. This advanced capability significantly boosts the model's proficiency in interpreting and managing dynamic graph data. The unique methodologies employed allow L-GAT to achieve outstanding performance in tasks that involve intricate graph structures.

### 4.2.1 Temporal Features Aggregation

At the outset of the L-GAT model, a critical procedure involves preprocessing the raw data features, particularly focusing on compressing the time-series data features via the LSTM algorithm. The LSTM is adept at evaluating information across all input time points, with the ability to selectively preserve or eliminate data through its distinctive gating mechanisms. Consequently, only essential information is maintained, while non-essential data is discarded. In this model, the resultant compressed data features are encapsulated into a compact feature representation at time $t$, denoted as $h_t$. This streamlined representation retains all pertinent historical information, forming the foundation for subsequent graph attention operations. Such a setup empowers the model to adeptly navigate and incorporate dynamic changes within time-series data into the broader analysis of the graph structure.

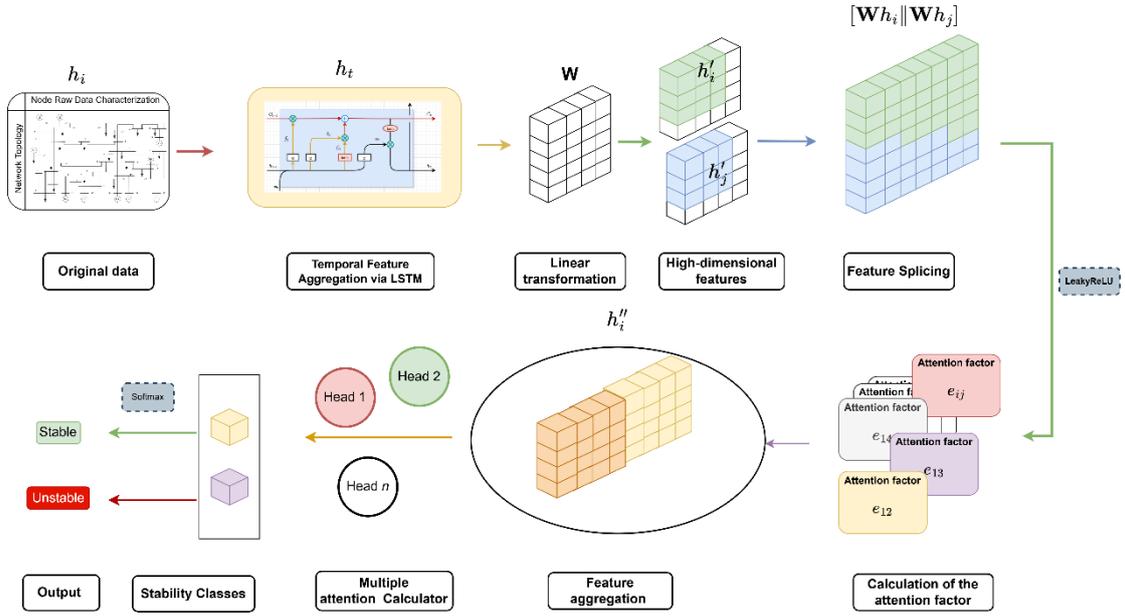

**Fig. 5.** Structure diagram of assessment L-GAT

**4.2.2 Linear Transformation of Node Features**

A pivotal operation within the L-GAT model is the linear transformation of each node's compressed output features into a new feature space. This transformation is executed by multiplying the features with a trainable weight matrix. This operation not only enhances but also normalizes the feature representations, preparing them for subsequent graph attention operations. Through this process, the model is optimally positioned to capture and process complex inter-node relationships effectively. The linear transformation can be mathematically represented as:

$$h'_i = \mathbf{W} h_i \tag{15}$$

where $h_i$ and $h'_i$ are respectively the original feature vector and its transformed feature vector of node $i$; $\mathbf{W}$ is the learnable weight matrix.

**4.2.3 Calculation of Attention Coefficients**

In L-GAT, the calculation of attention coefficients is a core step, as it determines the importance of each neighbor when aggregating the features of neighboring nodes.

$$e_{ij} = a(\mathbf{W}\vec{h}_i, \mathbf{W}\vec{h}_j) \tag{16}$$

$$\alpha_{ij} = \text{softmax}_j(e_{ij}) = \frac{\exp(e_{ij})}{\sum_{k \in \mathcal{N}_i} \exp(e_{ik})}. \tag{17}$$

$$\alpha_{ij} = \frac{\exp\left(\text{LeakyReLU}\left(a^T[\mathbf{W}h_i \parallel \mathbf{W}h_j]\right)\right)}{\sum_{k \in \mathcal{N}_i} \exp\left(\text{LeakyReLU}\left(a^T[\mathbf{W}h_i \parallel \mathbf{W}h_k]\right)\right)} \tag{18}$$

In this process, $a$ is a learnable attention vector used to weigh the feature importance between node $i$ and node $j$, $\parallel$ represents the concatenation operation, LeakyReLU(·) is the activation function, which introduces non-linearity, $\alpha_{ij}$ is the normalized attention weight from node $j$ to node $i$, ensuring that the sum of the weights of all neighbors equals 1.

### 4.2.4 Feature Aggregation

Using the computed attention coefficients, the new feature of node $i$ is the weighted sum of the features of all its neighbors:

$$h_i = \sigma\left(\sum_{j \in N_i} \alpha_{ij} \mathbf{W} h_j\right) \tag{19}$$

where $\sigma$ is the activation function, such as Sigmoid, used to further process the weighted aggregated features.

### 4.2.5 Multi-Head Attention Aggregation and Averaging

To improve the model's performance and generalization ability, L-GAT typically employs a multi-head attention mechanism. In this mechanism, the aforementioned process is repeated multiple times, each time using a different set of parameters (different $\mathbf{W}$ and $h$).

$$h_i' = \prod_{k=1}^{K} \sigma\left(\sum_{j \in \mathcal{N}_i} \alpha_{ij}^k \mathbf{W}^k h_j\right) \tag{20}$$

$$h_i' = \sigma\left(\frac{1}{K} \sum_{k=1}^{K} \sum_{j \in \mathcal{N}_i} \alpha_{ij}^k \mathbf{W}^k h_j\right) \tag{21}$$

Eq. (20) is typically used in the intermediate layers of the model to increase its expressive capacity. During training, the outputs from these attention heads can be concatenated, and in the final layer of the model, averaging is applied. Eq. (21) is generally used in the final layer or when smoother outputs are needed. It helps enhance the model's generalization ability and prevents overfitting by averaging the results from different attention heads, promoting stability and broader applicability. These formulas collectively define the fundamental operation flow of the L-GAT, enabling the model to effectively learn features of nodes in a graph while considering the strength and complexity of the relationships between nodes. In this way, L-GAT is able to adapt to various graph structures, providing robust support for a wide range of graph data analysis tasks. In addition, the hyperparameters of the L-GAT used in this paper are shown in **Table 1**.

**Table 1**. L-GAT hyperparameter settings

| Hyperparameter | Model Setting |
| --- | --- |
| Optimizer | Adam |
| Learning Rate | 0.0005 |
| Heads | 4 |
| Feature dimension/header | 32 |
| Batch Size | 64 |
| Iterations | 240 |
| Loss Function | Sparse Categorical Cross-entropy |

# 5 Proposed Methodology

To effectively perform STVSA under adversarial conditions, this study proposes a robust methodology based on L-GAT specifically tailored for real-time stability assessment of power systems, which is shown in Fig. 6. Recognizing the diverse nature of cyber-attacks, the method strategically employs mixed adversarial training, incorporating both WB-CAAs and BB-CAAs, to significantly enhance model resilience.

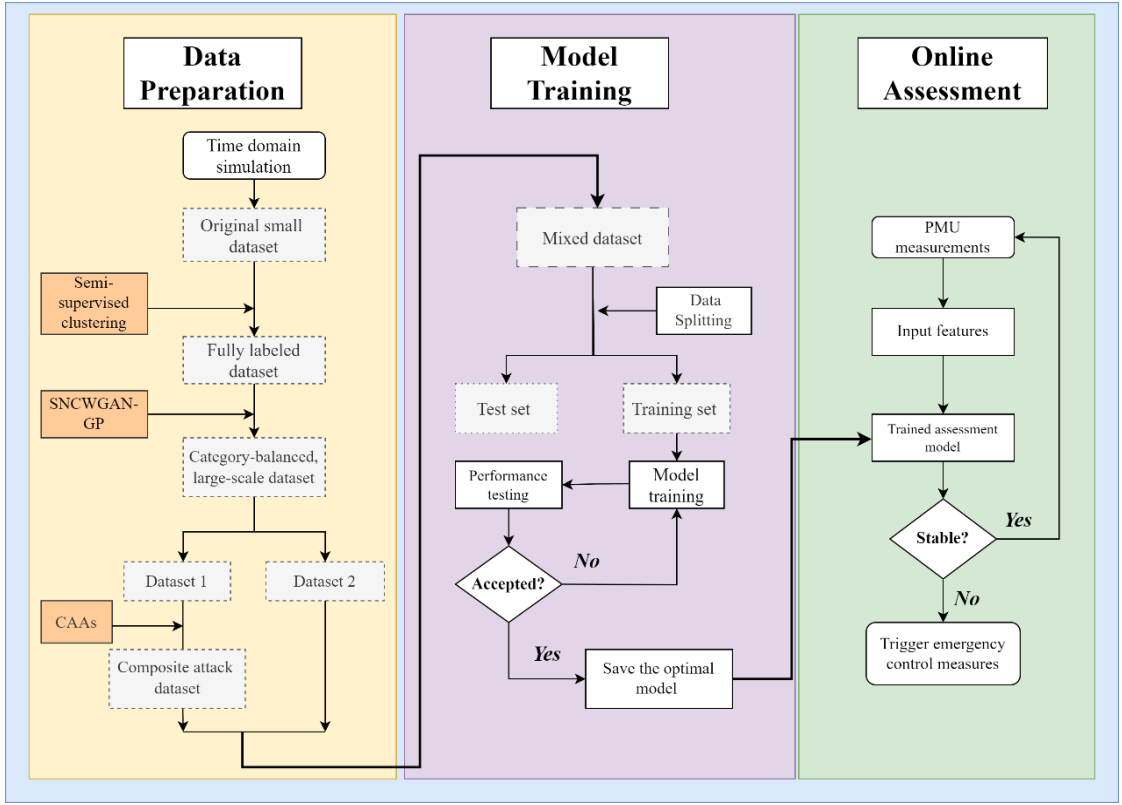

**Fig. 6.** Flowchart of the proposed L-GAT based STVSA methodology

As illustrated in Fig. 6, the methodology comprises three primary stages: data preparation, robust model training, and online assessment. This structured approach ensures efficient real-time data processing and robust responsiveness, thereby significantly strengthening the defense capabilities and operational stability of power systems against cyber threats.

## 5.1 Data Preparation

Due to the lack of a unified criterion for STVSA, this study utilizes a semi-supervised clustering method to resolve ambiguous sample labels at clustering boundaries, thus obtaining a fully labeled initial dataset. To mitigate class imbalance and expand data availability, the dataset undergoes data augmentation procedures described in Section 3, balancing and enlarging it effectively. To bolster model robustness comprehensively against potential cyber threats, the study generates adversarial samples using the FGSM. Critically, the study blends adversarial samples produced from WB-CAA and BB-CAA scenarios with clean data, thus forming a mixed dataset utilized for comprehensive adversarial training. This mixed training dataset ensures the model attains robust defensive capabilities capable of withstanding diverse, realistic cyber-attacks. Additionally, since the L-GAT model requires network topology data, accurate representation in graph form—where nodes symbolize dataset entities interconnected through edges defined by an adjacency matrix—is essential for effective graph neural network processing.

**5.2 Model Training**

The robust training process involves iterative forward and backward propagation: computing predictions through forward propagation, followed by calculating loss function gradients via back-propagation to iteratively refine model parameters. During training, critical hyperparameters such as learning rate, batch size, number of layers, and the number of attention heads are systematically optimized based on performance metrics. A suitable optimizer is carefully selected and employed to iteratively minimize the loss function by updating the parameters associated with the attention mechanism effectively. Upon reaching optimal model performance, configurations and corresponding hyperparameters are preserved for practical deployment.

**5.3 Online Assessment**

Once deployed, the robustly trained model continuously acquires real-time PMU data to monitor the power system's stability state. Upon detecting large disturbs, the system immediately evaluates stability by analyzing post-fault trajectory data. If instability is detected, emergency control measures are triggered promptly to mitigate potential risks. Conversely, if stability is confirmed, the system continues monitoring without manual intervention. This automated response mechanism enhances the system's operational resilience, providing reliable, real-time stability evaluation in diverse cyber-attack scenarios.

# 6 Case Analysis

All experiments were conducted using the IEEE 39-bus test system, with simulations run through PSD-BPA. The L-GAT model and all simulations, including adversarial and DoS attack scenarios, were implemented using Python, chosen for its robust library support and flexibility. Experiments utilized a desktop with an Intel Core i5-12600KF CPU, 32 GB RAM, and an RTX 4060 Ti GPU. The hyperparameters were finely tuned through trial and error to determine the optimal settings for the best experimental results, as detailed in Section 4 and listed in Table 2 for FGSM settings. All results were validated using 5-fold cross-validation to ensure reproducibility and accuracy [29].

**Table 2**. FGSM hyperparameter settings

| Hyperparameter | Specific Setting |
|---|---|
| Perturbation Strength | 0.04 |
| Optimizer | Adam |
| Learning Rate | 0.0001 |
| Threshold | 0.5 |
| Batch Size | 128 |
| Iterations | 1000 |
| Loss Function | Cross-Entropy Loss |

### 6.1 Dataset Generation

In this study, various power system operating conditions such as load ratios, induction motor load factors, and fault locations are simulated to generate an initial dataset of 3,400 samples. This dataset includes normalized physical quantities like node voltages, generator node outputs (both active and reactive power), consumption at load nodes, and power flow between adjacent nodes. A semi-supervised clustering algorithm is subsequently applied to classify boundary samples, yielding a fully labeled dataset. The dataset is then expanded to 12,000 samples using the SNCWGAN-GP method and split into training and testing sets at a 4:1 ratio for further analysis. It is worth noting that Gaussian white noise is added to the expanded test dataset to evaluate the impact of noise on model performance. Additionally, adversarial samples are created using the FGSM algorithm and DoS attack simulation methods to mimic the data state following a cyber-attack.

During data augmentation, the SNCWGAN-GP model generates new augmented samples based on the learned distribution of real data samples. At a probability distribution of 0.5 for both real and fake samples, the generated samples become indistinguishable to the discriminator. The training dynamics, illustrated in Fig. 7, show the generator's loss experiencing intense oscillations before stabilizing into convergent fluctuations, while the discriminator's loss progressively converges towards zero. This pattern indicates that the model has effectively learned the inherent features of the data through adversarial training. By continuously optimizing the parameters via adversarial training, the model succeeds in producing high-quality adversarial samples realistic enough to deceive the discriminator.

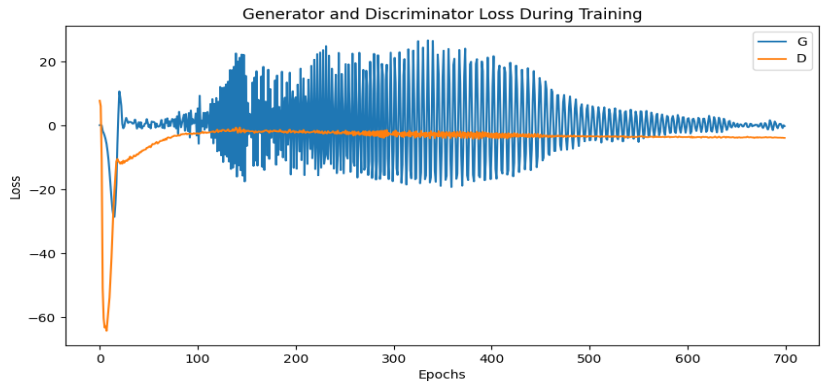

**Fig. 7.** Loss function of SNCWGAN-GP model in training

The performance of the SNCWGAN-GP model is quantitatively compared with that of the traditional LSGAN model using widely recognized metrics such as Wasserstein Distance (WD), Maximum Mean Discrepancy (MMD), and Fréchet Inception Distance (FID). These metrics assess the similarity and diversity of the generated data between the two models, with lower values indicating superior performance. The comparison results are presented in Table 3.

Table 3. Impact of cyber-attacks on different evaluation models

| Evaluation metrics | WD | MMD | FID |
|---|---|---|---|
| SNCWGAN-GP | 6.67 | 0.14 | 6.45 |
| LSGAN | 8.72 | 0.21 | 7.07 |

As shown in **Table 3**, on the high-dimensional dataset used in this study, the proposed method outperforms the traditional LSGAN algorithm across all three metrics. This indicates that SNCWGAN-GP exhibits better performance in generating complex data.

**6.2 Performance Analysis of STVSA under Cyber-attacks**

To rigorously evaluate the resilience of STVSA models against composite cyber-attacks, this study benchmarks five representative algorithms: L-GAT, Transformer, LSTM, Support Vector Machines (SVM), and Decision Trees (DT). The scenario-specific testing (WB-CAA vs. BB-CAA) is designed to assess the model's robustness under distinct threat scenarios, rather than to directly compare white-box and black-box attacks. By exposing the model to both attack types during training, we aim to enhance its generalization ability across a variety of adversarial conditions. Performance metrics under WB-CAA and BB-CAA scenarios are systematically compared in Tables 4 and 5, respectively.

Table 4. Performance of different models under WB-CAAs

| Algorithm | ACC (%) | AUC | MCC | F1-score | Epochs |
|---|---|---|---|---|---|
| L-GAT | 96.66% | 0.9531 | 0.9647 | 0.9622 | 240 |
| Transformer | 95.38% | 0.9596 | 0.9543 | 0.9552 | 300 |
| LSTM | 93.67% | 0.9232 | 0.9377 | 0.9268 | 200 |
| DT | 87.18% | 0.8767 | 0.8605 | 0.8826 | \ |
| SVM | 87.52% | 0.8678 | 0.8636 | 0.8694 | 100 |

As shown in Table 4, the proposed model exhibits better performance under WB-CAAs than the other alternatives. To be specific, the proposed L-GAT achieves 96.66% accuracy, outperforming other models by 1.28–9.48 percentage points. Despite adversarial perturbations causing metric declines, L-GAT maintains compliance with engineering requirements (ACC >95%). In contrast, shallow models (DT/SVM) exhibit severe performance degradation (ACC <88%), highlighting their vulnerability to compound attacks.

Table 5. Performance of different models under BB-CAAs

| Algorithm | ACC (%) | AUC | MCC | F1-score | Epochs |
|---|---|---|---|---|---|
| L-GAT | 94.13% | 0.9358 | 0.9439 | 0.9473 | 270 |
| Transformer | 92.58% | 0.9296 | 0.9243 | 0.9252 | 320 |
| LSTM | 90.27% | 0.9032 | 0.9077 | 0.9168 | 240 |
| DT | 83.18% | 0.8367 | 0.8205 | 0.8226 | \ |
| SVM | 82.82% | 0.8078 | 0.8236 | 0.8094 | 130 |

Table 5 summarizes model performance under BB-CAAs, where attackers cannot access internal model parameters. All models exhibit performance degradation due to incomplete adversarial information during training. However, the proposed L-GAT shows remarkable resilience, with only a 2.53% accuracy drop (94.13% vs. 96.66% in white-box scenarios). In contrast, shallow models (e.g., SVM and DT) suffer more severe accuracy losses (>4%), revealing their limitations in handling adversarial uncertainty, aligning with the no-free-lunch theorem—complex attack patterns necessitate deep architectures for generalization. These results validate the L-GAT's superiority in adversarial environments with partial observability, a critical challenge in power system cybersecurity.

To rigorously illustrate the impact of different cyber-attack scenarios on the robustness of STVSA models, test accuracy was chosen as the primary evaluation metric. The comparative performance of different models under WB-CAAs and BB-CAAs is intuitively presented in Fig. 8.

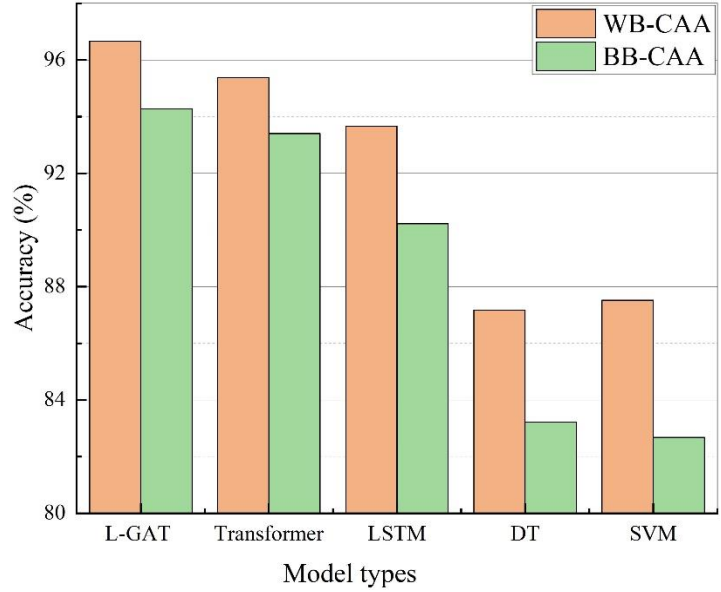

**Fig. 8.** Performance of the model under different attack scenarios

As observed from Fig. 8, models trained with adversarial samples generated under WB-CAAs demonstrate superior robustness compared to those trained under BB-CAAs. This advantage arises because white-box attackers possess comprehensive knowledge of the model's internal structure and parameters, enabling the creation of more precise and potent adversarial attacks. Consequently, models trained under WB-CAAs experience broader exposure to potential vulnerabilities, enhancing their defense against both known and unknown attack strategies, including black-box variants. Conversely, models trained only with BB-CAA samples lack comprehensive vulnerability coverage, limiting their protective effectiveness. In summary, adversarial training significantly enhances model resilience, particularly for deep learning approaches, effectively meeting the robustness criteria required for engineering applications.

**6.3 Performance Analysis of STVSA under Varying Noise Environments**

In real-world scenarios, PMU measurements inevitably encounter measurement noise [22]. To accurately assess the resilience of STVSA models in noisy operational environments, Gaussian white noise at signal-to-noise ratios (SNR) of 50 dB, 40 dB, and 30 dB was introduced into the test datasets [30]. The impact of these noise levels on model performance was evaluated through four metrics: ACC, AUC, MCC, and F1-score. Results of this comprehensive analysis are summarized in Table 6.

**Table 6.** STVSA performance under different noise environments

| Noise Environment | ACC (%) | AUC | MCC | F1-score |
|---|---|---|---|---|
| No Noise | 99.32% | 0.9925 | 0.9849 | 0.9884 |
| 50dB | 98.41% | 0.9869 | 0.9747 | 0.9623 |
| 40dB | 92.49% | 0.9243 | 0.9256 | 0.9215 |
| 30dB | 85.28% | 0.8481 | 0.8544 | 0.8477 |

From Table 6, it can be clearly seen that increasing noise intensity significantly degrades the model's performance. While the model exhibits optimal performance in noise-free conditions, its accuracy and other metrics gradually deteriorate with the introduction of stronger noise environments (50 dB, 40 dB, and 30 dB). Notably, at the lowest SNR of 30 dB, the model performance suffers the most severe degradation, underscoring the compounded destructive effects of cyber-attacks in noisy operational contexts. These results highlight the critical necessity for developing more robust evaluation models capable of maintaining stability and reliability even under adverse conditions.

### 6.4 Impact of Different Observation Time Windows on Evaluation Performance

The observation time window is a crucial parameter influencing the effectiveness of STVSA in power systems. Thus, an evaluation of the proposed model's performance across varying observation windows was conducted, comparing it against Transformer, LSTM, SVM, and DT-based models. The assessment utilized testing accuracy as a representative metric across datasets captured at 0.03 s, 0.06 s, 0.09 s, and 0.12 s after fault inception. The comparative results are illustrated in Fig. 9.

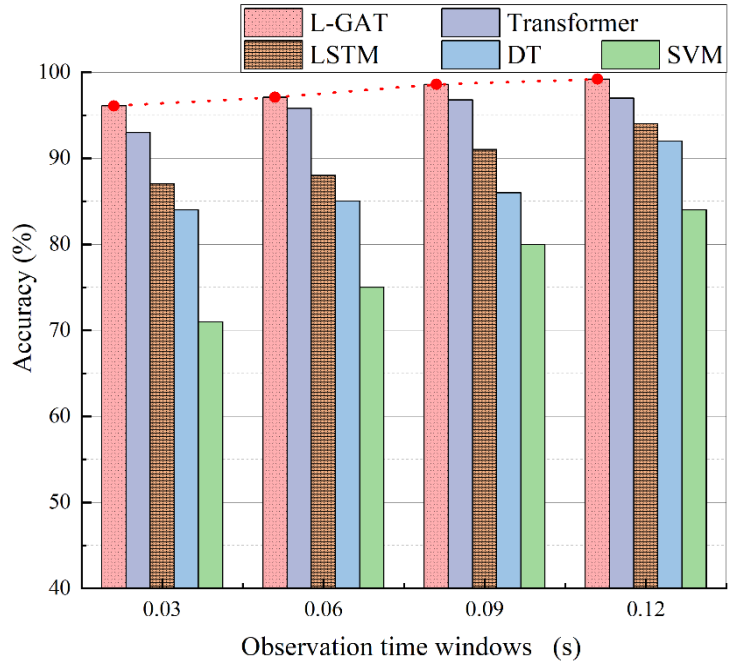

**Fig. 9.** Accuracy comparison across different observation time windows

As demonstrated in Fig. 9, the proposed L-GAT model consistently outperforms benchmark methods (Transformer, LSTM, SVM, and DT) across all observation time windows, clearly demonstrating its distinct advantage. Notably, an increase in observation time window length correlates positively with improved accuracy for all models, suggesting that larger sample sizes contribute to more accurate STVSA. Remarkably, even at the shortest observation window of 0.03 s—a critical threshold for real-time grid stability monitoring under rapid transient conditions—the proposed L-GAT model sustains accuracy above 95%, underscoring its robust capability to handle extreme system states.

# 7 Conclusions

Amidst the advancements of Industry 4.0, the digital integration of power systems into cyber-physical frameworks has increased their vulnerability to sophisticated cyber threats. To address this challenge, this study proposes an innovative AI-based control framework leveraging adversarial deep learning to enhance STVSA in power systems. Our primary motivation was to strengthen existing methods against adversarial cyber-attacks, thereby ensuring the reliability and stability of power system operations. The main conclusions of this work are as follows:

1) The developed L-GAT based model, which integrates LSTM networks with GAT, successfully captures dynamic relationships between time-series data and electrical grid topology. This integration allows for the simultaneous optimization of temporal and spatial analyses in STVSA, providing a robust methodological foundation for dynamic security assessments in power systems.
2) This study also investigates the model's resilience under composite cyber-attack scenarios, thoroughly analyzing composite adversarial attacks and their impacts. Despite these challenging conditions, the proposed adversarial deep learning-based approach demonstrates its effective and superiority in complex attack environments.
3) Building on existing CWGAN-GP based data augmentation techniques, this study improves its training stability and convergence by normalizing its spectral norm. This method not only improves data quality and the model's generalization ability but also enhances its robustness in adversarial testing scenarios.

This study achieves notable progress in enhancing STVSA resilience against composite cyber-attacks, several areas still require further exploration. While our experimental design ensures consistent perturbation strength across white-box and black-box attacks, the inherent differences in attack generation mechanisms may influence the evaluation results. Future work could explore more diverse attack generation strategies, such as using multiple surrogate models for black-box attacks, to further enhance the generalizability of the findings. Additionally, exploring scalability for more complex power systems and incorporating real-time data analytics will enable AI-driven models to dynamically adapt to evolving threats, advancing resilient control strategies for power system cybersecurity.

# Acknowledgment

This work was supported by the National Natural Science Foundation of China (No. 52377081) and the Natural Science Foundation of Jilin Province (No. YDZJ202101ZYTS149). Special thanks are extended to Mr. Chong Ma for his invaluable contributions and expert guidance during the revision phase of this manuscript, which significantly elevated the scholarly quality of the work.

# References

[1] Ma CYT, Yau DKY, Rao NSV. Scalable Solutions of Markov Games for Smart-Grid Infrastructure Protection. IEEE Trans Smart Grid 2013;4:47–55.

[2] Liang H, Li M, Chen Y, Jiang L, Xie Z, Yang T. Establishing Trusted I/O Paths for SGX Client Systems With Aurora. IEEE TransInformForensic Secur 2020;15:1589–600.

[3] Ten C-W, Yamashita K, Yang Z, Vasilakos AV, Ginter A. Impact Assessment of Hypothesized Cyberattacks on Interconnected Bulk Power Systems. IEEE Trans Smart Grid 2018;9:4405–25.

[4] Diao X, Zhao Y, Smidts C, Vaddi PK, Li R, Lei H, et al. Dynamic probabilistic risk assessment for electric grid cybersecurity. Reliability Engineering & System Safety 2024;241:109699.

[5] Liu M, Teng F, Zhang Z, Ge P, Deng R, Sun M, et al. Enhancing Cyber-Resiliency of DER-based SmartGrid: A Survey. IEEE Trans Smart Grid 2024;15:4998–5030.

[6] Zadkhast P, Jatskevich J, Vaahedi E. A Multi-Decomposition Approach for Accelerated Time-Domain Simulation of Transient Stability Problems. IEEE Trans Power Syst 2015;30:2301–11.

[7] Li Y, Zhang S, Li Y, Cao J, Jia S. PMU Measurements-Based Short-Term Voltage Stability Assessment of Power Systems via Deep Transfer Learning. IEEE Trans Instrum Meas 2023;72:1–11.

[8] Xu Y, Zhang R, Zhao J, Dong ZY, Wang D, Yang H, et al. Assessing Short-Term Voltage Stability of Electric Power Systems by a Hierarchical Intelligent System. IEEE Trans Neural Netw Learning Syst 2016;27:1686–96.

[9] Liu Y, Sun K, Yao R, Wang B. Power System Time Domain Simulation Using a Differential Transformation Method. IEEE Trans Power Syst 2019;34:3739–48.

[10] Khaitan SK, McCalley JD, Chen Q. Multifrontal Solver for Online Power System Time-Domain Simulation. IEEE Trans Power Syst 2008;23:1727–37.

[11] Potamianakis EG, Vournas CD. Short-Term Voltage Instability: Effects on Synchronous and Induction Machines. IEEE Trans Power Syst 2006;21:791–8.

[12] Lv Z, Wang B, Guo Q, Zhao H, Wang Z, Sun H. Short-Term Voltage Stability Assessment Based on Heterogeneous Edge-Integrated Graph Attention Network. IEEE Trans Power Syst 2024:1–13.

[13] Zhu L, Hill DJ, Lu C. Intelligent Short-Term Voltage Stability Assessment via Spatial Attention Rectified RNN Learning. IEEE Trans Ind Inf 2021;17:7005–16.

[14] Wang G, Zhang Z, Bian Z, Xu Z. A short-term voltage stability online prediction method based on graph convolutional networks and long short-term memory networks. International Journal of Electrical Power & Energy Systems 2021;127:106647.

[15] Luo Y, Lu C, Zhu L, Song J. Data-driven short-term voltage stability assessment based on spatial-temporal graph convolutional network. International Journal of Electrical Power & Energy Systems 2021;130:106753.

[16] Li Y, Ma W, Li Y, Li S, Chen Z, Shahidehpour M. Enhancing cyber-resilience in integrated energy system scheduling with demand response using deep reinforcement learning. Applied Energy 2025;379:124831.

[17] Qu Z, Dong Y, Li Y, Song S, Jiang T, Li M, et al. Localization of dummy data injection attacks in power systems considering incomplete topological information: A spatio-temporal graph wavelet convolutional neural network approach. Applied Energy 2024;360:122736.

[18] Ren C, Xu Y. A Universal Defense Strategy for Data-Driven Power System Stability Assessment Models Under Adversarial Examples. IEEE Internet Things J 2023;10:7568–76.

[19] Kong X, Lu Z, Guo X, Zhang J, Li H. Resilience Evaluation of Cyber-Physical Power System Considering Cyber Attacks. IEEE Trans Rel 2024;73:245–56.

[20] Farajzadeh-Zanjani M, Hallaji E, Razavi-Far R, Saif M. Generative-Adversarial Class-Imbalance Learning for Classifying Cyber-Attacks and Faults - A Cyber-Physical Power System. IEEE Trans Dependable and Secure Comput 2022;19:4068–81.

[21] Rbah Y, Mahfoudi M, Balboul Y, Fattah M, Mazer S, Elbekkali M, et al. Machine Learning and Deep Learning Methods for Intrusion Detection Systems in IoMT: A survey. 2022 2nd International Conference on Innovative Research in Applied Science, Engineering and Technology (IRASET), Meknes, Morocco: IEEE; 2022, p. 1–9.

[22] Golpîra H, Francois B. Artificial intelligence-based approach for islanding detection in cyber-physical power systems. Chaos, Solitons & Fractals 2024;185:115165.

[23] Xiong Z, Xu H, Li W, Cai Z. Multi-Source Adversarial Sample Attack on Autonomous Vehicles. IEEE Trans Veh Technol 2021;70:2822–35.

[24] Wang Y, Liu J, Chang X, Wang J, Rodríguez RJ. AB-FGSM: AdaBelief optimizer and FGSM-based approach to generate adversarial examples. Journal of Information Security and Applications 2022;68:103227.

[25] Liu M, Zhang Z, Chen Y, Ge J, Zhao N. Adversarial Attack and Defense on Deep Learning for Air Transportation Communication Jamming. IEEE Trans Intell Transport Syst 2024;25:973–86.

[26] Zhang D, Jin X, Su H. Event-Triggered Control Systems Under Stochastic Pulsing Denial-of-Service Attacks. IEEE Trans Automat Contr 2024;69:4013–20.

[27] Li Y, Zhang M, Chen C. A Deep-Learning intelligent system incorporating data augmentation for Short-Term voltage stability assessment of power systems. Applied Energy 2022;308:118347.

[28] Veličković P, Cucurull G, Casanova A, Romero A, Liò P, Bengio Y. Graph Attention Networks 2018. https://doi.org/10.48550/arXiv.1710.10903.

[29] Liu Z, Yang Y. Selection of uncertain differential equations using cross validation. Chaos, Solitons & Fractals 2021;148:111049.

[30] Li Y, Cao J, Xu Y, Zhu L, Dong ZY. Deep learning based on Transformer architecture for power system short-term voltage stability assessment with class imbalance. Renewable and Sustainable Energy Reviews 2024;189:113913.